\documentclass[aps,prl,twocolumn]{revtex4}
\usepackage{graphicx}
\usepackage{dcolumn}
\usepackage{amsmath}
\usepackage{amssymb}
\usepackage{subfigure,amsmath,verbatim,moreverb,bm}
\def\be{\begin{equation}}
\def\ee{\end{equation}}
\def\ber{\begin{eqnarray}}
\def\eer{\end{eqnarray}}

\def\rv{{\bf r}}
\def\Rv{{\bf R}}
\def\sv{{\bf s}}
\def\fv{{\bf f}}

\begin{document}
\title{Density functional theory for strongly interacting electrons}
\author{Paola Gori-Giorgi,$^{1,2}$ Michael Seidl,$^3$ and G. Vignale$^4$}
\affiliation{$^1$Laboratoire de Chimie Th\'{e}orique, CNRS,
Universit\'{e} Pierre et Marie Curie, 4 Place Jussieu, 75252 Paris, France \\
$^2$Afd. Theoretische Chemie, Vrije Universiteit, De Boelelaan 1083, 1081   HV Amsterdam, The Netherlands\\
$^3$Institute of Theoretical Physics,
University of Regensburg, 93040 Regensburg, Germany \\
$^4$Department of Physics and Astronomy,
University of Missouri, Columbia, Missouri 65211, USA}
\date{\today}
\begin{abstract}
We present an alternative to the Kohn-Sham formulation of density functional theory for the ground-state properties of strongly interacting electronic systems.   The idea is to start from the limit of zero kinetic energy and systematically expand the universal energy functional of the density in powers of a ``coupling constant'' that controls the magnitude of the kinetic energy.  The problem of minimizing the energy is reduced to the solution of a  strictly correlated electron problem in the presence of an effective potential, which plays in our theory the same role that the Kohn-Sham potential plays in the traditional formulation. We discuss several schemes for approximating the energy functional, and report preliminary results for low-density quantum dots. 
\end{abstract}
\maketitle
Electronic systems are classified as weakly or strongly correlated depending on whether the potential energy is much smaller or much larger than the kinetic energy.  In the first case the system is well described in terms of independent one-electron orbitals.  In the second case it tends to crystallize in a ``Wigner molecule" -- a state in which the position of a single electron determines the position of all the others, even as any given electron is distributed in space according to the average density of the system.   We will refer to this limit as the {\it strictly correlated electron} (SCE) limit. 

Which of the two descriptions is more accurate for a given physical system  depends on the ratio of the average inter-electron distance $\ell$ to the effective Bohr radius $a^*=\hbar^2/m^* {e^*}^2$ (including appropriately screened charge $e^*=e/\epsilon$ and effective mass $m^*$): the strongly correlated regime occurs when $\ell/a^* \gg 1$ and the weakly correlated regime when $\ell/a^* \ll1$.   For example,
a system of $N$ electrons trapped in the parabolic potential of a semiconductor quantum dot (``artificial atom")  will enter the strongly correlated regime when the confinement length becomes larger than the effective Bohr radius.  Similarly, a chain of hydrogen atoms becomes more strongly correlated as the distance between the protons increases.  Furthermore all systems tend to become more strongly correlated as the effective dimensionality is reduced. From a traditional quantum chemistry point of view, strongly correlated systems need very many (billions) of Slater determinants for a reasonable description of their physics. In other words, they are characterized by natural occupation numbers that are all close to zero. 

Strongly correlated systems pose a severe challenge to any many-body theory, because the electron-electron interaction cannot be treated perturbatively.  In this paper we focus on the treatment of strongly correlated systems within the framework of density functional theory (DFT).   DFT offers in principle a way to deal uniformly with both weakly and strongly interacting systems.  In the original formulation of Hohenberg and Kohn (HK) \cite{HohKoh-PR-64} the ground-state density  and energy are obtained by minimizing with respect to the density $\rho(\rv)$ the energy functional
\be\label{EnergyFunctional}
E[\rho] = F[\rho]+\int d\rv\, v_{\rm ext}(\rv)\,\rho(\rv),
\ee
where $v_{\rm ext}(\rv)$ is the external potential  and $F[\rho]$ is a universal functional of the density, defined as the expectation value of the internal energy (kinetic energy $\hat T$ plus electron-electron interaction energy $\hat V_{ee}$) in the  unique ground-state wave function that yields the density $\rho(\rv)$.  

It is standard practice to  carry out the minimization of Eq.~(\ref{EnergyFunctional}) by the Kohn-Sham (KS) method \cite{KohSha-PR-65}.  This method introduces the functional $T_s[\rho]$ by minimizing the expectation value of  $\hat T$  alone over all wavefunctions yielding the given $\rho$, and thus introducing a reference non-interacting system.  The remaining parts of the exact energy functional are approximated.  This works very well when the kinetic energy dominates, but runs into difficulties as the system becomes more strongly interacting.  Indeed, a proper treatment of strong correlation is considered one of the two major problems facing DFT today (the other being a proper inclusion of van der Waals interactions).   

When the electron-electron repulsion dominates over the kinetic energy it could be much better to do the opposite: define a model system in which one minimizes $\hat V_{ee}$ alone over all wavefunctions yielding the given $\rho$, and approximate the remaining terms.  This is precisely the approach we take in this paper. 
Thus, we present a rigorous formulation of DFT that is alternative and complementary to the traditional KS-DFT approach in the sense that our reference system is a strictly-correlated system rather than the non-interacting KS system. This approach should be more suitable to treat systems whose density is not dominated by the quantum mechanical shells, but by incipient Wigner-crystallization effects. 

To formulate the problem precisely we start with the standard many-electron hamiltonian
\be \label{model}
\hat H = \hat T+ \hat V_{ee}+\hat V_{\rm ext}
\ee
in which $\hat T = -\frac{1}{2m^*}\sum_i\nabla_i^2$ is the kinetic energy (we set $\hbar=1$ throughout),  $\hat V_{ee} = \sum_{i<j} \frac{e^2}{\epsilon |\rv_i-\rv_j|}$ is the electron-electron interaction,  and $\hat V_{\rm ext} = \sum_i v_{\rm ext}(\rv_i)$  is the external potential. The internal energy functional $F[\rho]$ is defined as \cite{Lev-PNAS-79}
\be
F [\rho] = \langle \Psi[\rho]|\hat T + \hat V_{ee}|\Psi[\rho]\rangle
\ee
where $\Psi[\rho]$ is the ground-state fermionic wave function uniquely associated with the density $\rho$. 

We further separate $F[\rho]$ into a strictly correlated contribution $V_{ee}^{\rm SCE}[\rho]$ defined as the interaction energy functional for a system with zero kinetic energy (i.e., the minimum of $\hat V_{ee}$ alone over all wavefunctions that yield the density $\rho$), and a remainder, which we call {\it kinetic-decorrelation energy functional} $E_{kd}[\rho]$:
\be \label{defEkc}
F [\rho] =V_{ee}^{\rm SCE} [\rho]+E_{kd}[\rho]\,.
\ee
Observe the close analogy between this separation and the conventional one involving the noninteracting kinetic energy, Hartree energy, and exchange-correlation energy. The functional $E_{kd}[\rho]$ corresponds to the expectation of the kinetic energy $\hat T$ plus the corrections to the expectation of $\hat V_{ee}$. In this Letter, we explain how to construct the functional $V_{ee}^{\rm SCE} [\rho]$ and its functional derivative, we present three possible approximations for $E_{kd}[\rho]$, and we report preliminary results for low-density quantum dots.
 
The functional $V_{ee}^{\rm SCE}[\rho]$ corresponds to the strong-interaction limit of the traditional adiabatic connection of DFT and was first addressed, in an approximate way, about 10 years ago \cite{Sei-PRA-99,SeiPerLev-PRA-99,SeiPerKur-PRA-00}. Only recently Seidl,  Gori-Giorgi  and Savin \cite{SeiGorSav-PRA-07} have found a general procedure for constructing the functional $V_{ee}^{\rm SCE}[\rho]$, as well as other corresponding observables such as the pair-correlation function \cite{GorSeiSav-PCCP-08}. Their construction can be viewed as the generalization 
of the Wigner-correlated regime to any given non-uniform smooth density $\rho(\rv)$. The key difference is that in the traditional Wigner regime (e.g., in Wigner atoms or molecules, or in the Wigner-crystal case), the electronic density is determined by the classical minimum of the hamiltonian without kinetic energy. Here, instead, we fix the density a priori, and for any given {\em smooth} (quantum mechanical) density we construct the corresponding $V_{ee}^{\rm SCE}[\rho]$. 

In the strong-interaction limit of DFT (no kinetic energy, but fixed given density $\rho$) the admissible configurations of $N$ electrons in $d$ dimensions are restricted to a $d$-dimensional subspace of the full $Nd$-dimensional configuration space \cite{SeiGorSav-PRA-07}.   We call this subspace $\Omega_0$.  A generic point of $\Omega_0$ has the form
\be
\Rv_{\Omega_0}(\sv) = (\fv_1(\sv),....,\fv_N(\sv))\,,
\ee
where $\sv$ is a $d$-dimensional vector that determines the position of, say, electron ``1", 
 and $\fv_i(\sv)$ ($i=1,...,N$, $\fv_1(\sv)=\sv$) are the {\it co-motion functions},  which determine the position of the $i$-th electron in terms of $\sv$.   The variable $\sv$ itself is distributed according to the normalized density $\tilde \rho(\sv) \equiv \rho(\sv)/N$.  The co-motion functions are implicit functionals of the density, determined by a set of differential equations that ensure the invariance of the density under the coordinate transformation $\sv \to \fv_i(\sv)$, i.e. $\rho(\fv_i(\sv))d\fv_i(\sv)=\rho(\sv)d\sv$ \cite{SeiGorSav-PRA-07}.  They play the same role in our theory as the Kohn-Sham orbitals in the conventional formulation of DFT.  In particular, the $\fv_i$ determine the functional $V_{ee}^{\rm SCE}[\rho]$ through the equation
\be
V_{ee}^{\rm SCE}[\rho]=\int d\sv\, \tilde \rho(\sv) \, \sum_{i<j}\frac{e^2}{\epsilon|\fv_i(\sv)-\fv_j(\sv)|}\,,
\ee
just as the Kohn-Sham orbitals determine the non-interacting kinetic energy.  Further, the total potential energy of a classical configuration  
\be
E_{pot}(\rv_1,...,\rv_N)= \sum_{i<j}\frac{e^2}{\epsilon|\rv_i -\rv_j|}+\sum_i v_{\rm SCE}[\rho](\rv_i)\,,
\ee
 where $v_{\rm SCE}[\rho](\rv)$ is the external potential associated with the density $\rho$ at zero kinetic energy, is constant on $\Omega_0$ \cite{SeiGorSav-PRA-07} and is expected to have a minimum with respect to variations perpendicular to $\Omega_0$, implying that its Hessian has $d$ eigenvectors with null eigenvalue and $Nd-d$ eigenvectors with positive eigenvalue at every point on $\Omega_0$ \cite{GorVigSei-JCTC-09}.
The co-motion functions $\fv_i$ have been constructed for a general spherical density \cite{SeiGorSav-PRA-07,GorVigSei-JCTC-09}, while the solution in the general case is still the object of our on-going work. Although the functional $V_{ee}^{\rm SCE}[\rho]$ depends on the density $\rho$ in an implicit way through the co-motion functions $\fv_i(\rv)$, its functional derivative with respect to $\rho(\rv)$ is
\be
\frac{\delta V_{ee}^{\rm SCE}[\rho]}{\delta \rho(\rv)}=-v_{\rm SCE}[\rho](\rv),
\ee
where the potential $v_{\rm SCE}[\rho](\rv)$ satisfies the classical equilibrium equation \cite{SeiGorSav-PRA-07}
\be
\nabla v_{\rm SCE}[\rho](\rv)=\sum_{i= 2}^N \frac{\rv-\fv_i(\rv)}{|\rv-\fv_i(\rv)|^3}.
\ee
Combining Eqs.~(\ref{defEkc}) and~(\ref{EnergyFunctional}) we see that the minimization of the energy leads to the variational condition
\be
\frac{\delta V_{ee}^{\rm SCE}[\rho]}{\delta\rho(\rv)}= -v_{\rm ext}(\rv) + \mu - \frac{\delta E_{kd}[\rho]}{\delta\rho(\rv)}\,,
\ee 
where $\mu$ is a constant (chemical potential).  Assuming that we know how to approximate $E_{kd}[\rho]$ and how to calculate its functional derivative,
\be
v_{kd}[\rho](\rv) \equiv  \frac{\delta E_{kd}[\rho]}{\delta\rho(\rv)},
\ee
we therefore reduce the energy minimization to the problem of solving a strictly correlated system  in an effective potential $v_{\rm SCE}[\rho](\rv)=v_{\rm ext}(\rv)+v_{kd}[\rho](\rv)$. An advantage with respect to the KS approach is that the co-motion functions $\fv_i(\rv)$ can be directly constructed from the density by integrating the differential equation $\rho(\fv_i(\sv))d\fv_i(\sv)=\rho(\sv)d\sv$ \cite{SeiGorSav-PRA-07}. Thus, the simplest way to solve the SCE equations is probably by directly minimizing the energy with respect to the density, using a proper basis set or a grid.

The central problem is to obtain an explicit expression for $E_{kd}[\rho]$.  Here we discuss three approximations: a ``first-order'' approximation, a proper generalization of the local-density approximation, and the combination of the two.
To define a ``first-order'' approximation, we parallel the standard adiabatic connection approach to the calculation of the exchange-correlation energy.   Namely, we introduce a fictitious hamiltonian
\be \label{AdiabaticConnection}
\hat H_\alpha = \alpha\, \hat T+ \hat V_{ee}+\hat V_\alpha
\ee
where $\alpha$ is a positive constant that takes values between $0$ and $1$  and $\hat V_\alpha = \sum_i v_\alpha(\rv_i)$ is an external potential chosen in such a way as to yield the desired ground-state density $\rho(\rv)$ for every value of $\alpha$.  Notice that, at variance with the standard approach, we make an adiabatic connection between the reference system (strictly correlated electrons at $\alpha=0$) and the physical system ($\alpha=1$) by gradually turning on $\hat{T}$ rather than $\hat{V}_{ee}$.  Making use of the Hellman-Feynman theorem it is easy to show that
\be\label{Hellman-Feynman}
E_{kd} [\rho] = \int_0^1 d\alpha \langle \Psi_\alpha[\rho]|\hat T|\Psi_\alpha[\rho]\rangle,
\ee
where $\Psi_\alpha[\rho]$ is the ground-state wave function associated with the density $\rho$ at coupling constant $\alpha$.  We have recently shown \cite{GorVigSei-JCTC-09} that the exact expansion of $T_\alpha[\rho] \equiv \langle \Psi_\alpha[\rho]|\hat T|\Psi_\alpha[\rho]\rangle $ for $\alpha \to 0$  can be obtained from a classical harmonic analysis, which yields
\be\label{TExpansion}
T_\alpha[\rho] =  \alpha^{-1/2}\,T_{\rm ZP}[\rho]+O(\alpha^0),
\ee
where $T_{\rm ZP}[\rho]$ is the kinetic energy associated with zero-point oscillations about the SCE solution. Inserting this expansion into Eq.~(\ref{Hellman-Feynman}) yields a ``first-order'', or zero-point (ZP) expression for $E_{kd} [\rho]$,
\be
\label{EkcZP}
E_{kd}^{\rm ZP} [\rho]=2\,T_{\rm ZP}[\rho]=\int d\sv\,\tilde \rho(\sv)\,\sum_{n=1}^{Nd-d} \frac{\omega_n(\sv)}{2},
\ee
where $\omega_n(\sv)$ are the $Nd-d$ zero-point frequencies around the degenerate SCE minimum \cite{GorVigSei-JCTC-09}. An example of calculation of $T_{\rm ZP}[\rho]$ for spherical atoms from He to Ne is reported in Ref.~\onlinecite{GorVigSei-JCTC-09}. While Eq.~(\ref{TExpansion})  is formally valid only in the limit $\alpha \to 0$ it is essential to appreciate that, in a physical sense, a small value of $\alpha$ is one for which $\alpha \ll  \ell/a^*$.  It follows that for a strongly correlated system, in which $\ell/a^*\gg 1$, the physical value of $\alpha=1$ is already in the strongly interacting regime. 

A simpler approximation is a generalization of LDA,
\be
\label{EkcLDA}
E_{kd}^{\rm LDA}[\rho]=\int d\rv\,\rho(\rv)\,\epsilon_{kd}(\rho(\rv)),
\ee
where $\epsilon_{kd}(\rho)$ is the kinetic-decorrelation energy of a uniform electron gas with density $\rho$, given by
\be
\label{eq_LDA}
\epsilon_{kd}(r_s)=t_s(r_s)+\epsilon_{xc}(r_s)-\frac{a_M}{r_s}.
\ee
Here $r_s$ is the density parameter, defined by  $r_s a^* =  (\rho B_d)^{-1/d}$, where $B_d$ is the ``volume" of the $d$-dimensional ball of unit radius \cite{ GiuVig-BOOK-05}.  The coefficient $a_M$ determines the Madelung energy. Notice, again, the analogy with the standard KS-LDA: in the Kohn-Sham formalism, $t_s$ is treated exactly via the functional $T_s[\rho]$, so that it is subtracted from the total energy of the electron gas. In our case, $V_{ee}^{\rm SCE}[\rho]$ is treated exactly via the construction of the co-motion functions, and so the corresponding value for the electron gas (the Madelung energy) is subtracted from its total energy. Another way to look at Eq.~(\ref{eq_LDA}) is that LDA is uniquely defined as the approximation that makes Eq.~(\ref{defEkc}) exact for a uniform density. As in standard KS theory, LDA has the advantage of being an explicit functional of the density, so that its functional derivative is easily calculated.

A third approximation can be obtained by combining the zero-point energy and the LDA,
\be
\label{EkcLDAZP}
E_{kd}^{\rm LDA-ZP}[\rho]=2\,T_{\rm ZP}[\rho]+\int d\rv\,\rho(\rv)\,\epsilon_{kd{\rm -ZP}}(\rho(\rv)),
\ee
where
\be
\label{eq_LDA-ZP}
\epsilon_{kd{\rm -ZP}}(r_s)=t_s(r_s)+\epsilon_{xc}(r_s)-\frac{a_M}{r_s}-\frac{a_{\rm ZP}}{r_s^{3/2}},
\ee
is the total energy of the uniform electron gas from which we have now subtracted also the zero-point low-density term. Again, the spirit is to define LDA as the approximation that makes Eq.~(\ref{EkcLDAZP}) exact for a uniform density. 

\begin{table}[t]
\begin{tabular}{cccccc}
\hline\hline
$\omega$                   &  KS-LDA   & SCE   & SCE-ZP & SCE-LDA   & SCE-ZP-LDA   \\
\hline                                                                         
$1.000 \times 10^{0}$      & 2.0       & 40.4  & 17.7   &  3.4      & 14.3   \\  
$1.667\times 10^{-1}$      & 2.4       & 32.7  & 11.2   & 4.8       & 14.9   \\
$5.393\times 10^{-2}$      & 1.6       & 27.1  & 8.0    & 5.5       & 14.1   \\
$2.368\times 10^{-2}$      & 0.1       & 23.0  & 6.1    & 5.8       & 13.1   \\
$7.285 \times 10^{-3}$     & 4.2       & 17.6  & 4.2    & 5.6       & 10.9   \\
$2.211\times 10^{-3}$      & 11.6      & 13.0  & 2.8    & 4.8       & 8.1   \\
$1.221\times 10^{-3}$      & 16.5      & 11.0  & 2.3    & 4.3       & 6.8   \\
$5.973\times 10^{-4}$      & 23.4      & 9.0   & 1.8    & 3.6       & 5.3   \\
$3.353 \times 10^{-4}$     & 29.7      & 7.6   & 1.5    & 3.1       & 4.2   \\
$2.408\times 10^{-4}$      & 33.6      & 6.9   & 1.4    & 2.8       & 3.6   \\

\hline\hline  
\end{tabular}
\caption{Relative \% errors on the total energy of a model 2D quantum dot consisting of two electrons confined in an harmonic potential $v_{\rm ext}(\rv)=\frac{1}{2}\omega^2 r^2$. Columns as follows: KS-LDA are the results for standard Kohn-Sham LDA, SCE are the results obtained by setting $E_{kd}[\rho]=0$ in Eq.~(\ref{defEkc}),  SCE-ZP are the results obtained from Eq.~(\ref{EkcZP}),  SCE-LDA are those obtained by using $E_{kd}^{\rm LDA}[\rho]$ of Eqs.~(\ref{EkcLDA})-(\ref{eq_LDA}), and SCE-ZP-LDA are those obtained with Eqs.~(\ref{EkcLDAZP})-(\ref{eq_LDA-ZP}). 
}
\label{tab_resN2}
\end{table}

As a preliminary test for our construction and approximations we have used a simple two-dimensional (2D) quantum-dot model consisting of two electrons confined in an harmonic potential $v_{\rm ext}(\rv)=\frac{1}{2}\omega^2 r^2$, with $\omega$ measured in effective Hartree$^*$. At this first stage we have used as inputs the exact densities of Ref.~\onlinecite{Tau-PAMG-94}. The 2D-LDA functional is from Ref.~\onlinecite{AttMorGorBac-PRL-02}. The \% errors on the total energy are reported in Table~\ref{tab_resN2}, where we compare, for different values of $\omega$, the standard Kohn-Sham LDA results (column KS-LDA) with those from the SCE construction with $E_{kd}[\rho]=0$ (SCE), those from the ``first-order'' approximation (SCE-ZP), those obtained by using $E_{kd}^{\rm LDA}[\rho]$ (SCE-LDA), and those from the combination of the two (SCE-ZP-LDA). We see that when the system is weakly correlated (large $\omega$) the KS-LDA result is superior, as the physics of the system is well captured by the non-interacting reference system. But as $\omega$ is lowered and the system becomes more correlated, the SCE construction with its approximations for $E_{kd}[\rho]$ becomes much more accurate than KS-LDA. The ``first-order'' (SCE-ZP) approximation gives the best results in the strongly-correlated regime (small $\omega$), but the simpler SCE-LDA result is also accurate over a broad range of $\omega$ values \footnote{The good performance of SCE-LDA also for large $\omega$ is due to the exactness of LDA for the non-interacting kinetic energy in a 2D harmonic trap, see M. Brack and B. van Zyl, Phys. Rev. Lett. {\bf 86}, 1574 (2001)}, reducing the KS-LDA errors of a factor $\sim 4$ up to $\sim 10$ as the system becomes more correlated. The combination of ``first-order'' and LDA (SCE-ZP-LDA), instead, is always worse than the simpler SCE-LDA approximation. This is very similar to what happens in standard KS-DFT when we combine the exact first-order approximation (which, in this case, is the exact exchange) with the LDA correlation energy: the results are worse than when using LDA for both exchange and correlation. The poor performance of KS-LDA for small $\omega$ is due to the fact that the single Slater determinant is a very bad approximation in this regime: exact-exchange yields much worse results, e.g., overestimating the total energy by $\sim 60\%$ at $\omega=2.211\times 10^{-3}$.

We note that the two functionals $V_{ee}^{\rm SCE}[\rho]$ and $T_{\rm ZP}[\rho]$ are formally independent of the statistics of the particles: that is because in the strongly correlated limit   the particles are distinguished by their relative positions, and exchange effects are in a first approximation negligible, entering, formally, at orders $e^{-\alpha^{-1/4}}$ (i.e., the order of magnitude of the overlap of different gaussians in the zero-point oscillations).  However, such exchange effects are in principle contained in $E_{kd}[\rho]$: its LDA approximation, for example, takes into account, in an approximate way, the fermionic nature of the system. 

In conclusion, we have presented a new formalism with high potential to treat strongly interacting systems such as low-density nanodevices.  A representative of the class of problems that can be tackled by this formalism is the calculation of the addition energy of  quantum dots \cite{Ash-N-96,TarAusHonHagKow-PRL-96}.  Aside from the practical importance of the problem (addition energies control the threshold potentials for one-electron transistors), experiments done in the low-density regime have revealed intriguing patterns \cite{ZhiAshPfeWes-PRL-97}, which are suggestive of Wigner-like correlations and have never been fully explained.  This kind of electronic structure calculations have been, so far, only accessible to wavefunction methods such as Quantum Monte Carlo \cite{GhoGucUmrUllBar-NP-06} and configuration interaction (only small dots) \cite{RonCavBelGol-JCP-06}. The KS approach is, in this context, only useful in the moderately correlated regime \cite{ReiMan-RMP-02}. 
Moreover, similarly to recently proposed first-order density-matrix energy functionals \cite{GriPerBae-JCP-05}, the functional $V_{ee}^{\rm SCE}[\rho]$ yields by definition the exact dissociation limit of the H$_2$ molecule, a typical case in which restricted KS calculations fail. The more challenging case of moderately stretched H chains will be tested in future work. 
 We believe that our approach will prove more effective than traditional KS for these kind of problems \footnote{Some similar formalism has been recently and independently discussed by Z.-F. Liu and K. Burke, arXiv:0907.2736.}, contingent on the development of an efficient algorithm to routinely solve the SCE equations. A first step in this direction is to reformulate $V_{ee}^{\rm SCE}[\rho]$ as a generalized mass transportation problem \cite{ButDepGor-PRA-XX}.

P.G.G.~was supported by ANR (07-BLAN-0272), and
G.V.~was supported by DOE Grant No.~DE-FG02-05ER46203.  P.G.G.~acknowledges discussions with K. Burke and A. Savin.

\end{document}